\documentclass[preprint,11pt]{JHEP} 
\usepackage{epsfig}
\usepackage{amsfonts,amssymb}

\newcommand{\al}{\ensuremath{\alpha}}
\newcommand{\ga}{\ensuremath{\gamma}}
\newcommand{\Ga}{\ensuremath{\Gamma}}
\newcommand{\de}{\ensuremath{\delta}}

\newcommand{\ep}{\ensuremath{\epsilon}}

\newcommand{\ka}{\ensuremath{\kappa}}
\newcommand{\la}{\ensuremath{\lambda}}
\newcommand{\om}{\ensuremath{\omega}}
\newcommand{\Om}{\ensuremath{\Omega}}
\newcommand{\p}{\ensuremath{\phi}}
\newcommand{\s}{\ensuremath{\sigma}}
\renewcommand{\S}{\ensuremath{\Sigma}}

\renewcommand{\th}{\ensuremath{\theta}}

\newcommand{\D}{\ensuremath{{\cal D}}}
\newcommand{\E}{\ensuremath{{\cal E}}}
\newcommand{\F}{\ensuremath{{\cal F}}}

\renewcommand{\L}{\ensuremath{{\cal L}}}

\newcommand{\N}{\ensuremath{{\cal N}}}
\renewcommand{\O}{\ensuremath{{\cal O}}}

\newcommand{\ra}{\ensuremath{\rightarrow}}
\newcommand{\del}{\ensuremath{\partial}}

\newcommand{\inv}{\ensuremath{^{-1}}}
\newcommand{\half}{\ensuremath{\frac{1}{2}}}

\newcommand{\quarter}{\ensuremath{\frac{1}{4}}}
\newcommand{\ul}{\ensuremath{\underline}}

\newcommand{\be}{\begin{equation}}
\newcommand{\ee}{\end{equation}}

\newcommand{\ba}{\begin{eqnarray}}
\newcommand{\ea}{\end{eqnarray}}

\newcommand\fverb{\setbox\pippobox=\hbox\bgroup\verb}
\newcommand\fverbdo{\egroup\medskip\noindent%
  \fbox{\unhbox\pippobox}\ }
\newcommand\fverbit{\egroup\item[\fbox{\unhbox\pippobox}]} \newbox\pippobox

\title{On the Eleven-Dimensional Origins of Polarized D0-branes}

\author{D. Brecher\\
  Centre for Particle Theory, Department of Mathematical Sciences,\\ 
  University of Durham, South Road, Durham DH1 3LE, United Kingdom.\\
  E-mail: \email{dominic.brecher@durham.ac.uk}}

\author{A. Chamblin\\
  Center for Theoretical Physics, Massachusetts Institute of
  Technology, \\ Bldg. 6-304, Cambridge, MA 02139, U.S.A. \\
  E-mail: \email{chamblin@mit.edu}}

\preprint{\hepth{0104046}\\DCPT/01/29\\MIT-CTP-3105}

\abstract{The worldvolume theory of a D0-brane contains a multiplet of fermions
which can couple to background spacetime fields.  This coupling
implies that a D0-brane may possess multipole moments with respect to
the various type IIA supergravity fields.  Different such polarization
states of the D0-brane will thus generate different long-range
supergravity fields, and the corresponding semi-classical supergravity
solutions will have different geometries.  In this paper, we
reconsider such solutions from an eleven-dimensional perspective.  We
thus begin by deriving the ``superpartners'' of the eleven-dimensional
graviton.  These superpartners are obtained by acting on the purely
bosonic solution with broken supersymmetries and, in theory, one can
obtain the full BPS supermultiplet of states.  When we dimensionally
reduce a polarized supergraviton along its direction of motion, we
recover a metric which describes a polarized D0-brane.  On the other
hand, if we compactify along the retarded null direction we obtain the
short distance, or ``near-horizon'', geometry of a polarized D0-brane,
which is related to finite $N$ Matrix theory.  The various dipole
moments in this case can only be defined once the eleven-dimensional
metric is ``regularized'' and, even then, they are formally infinite.
We argue, however, that this is to be expected in such a
non-asymptotically flat spacetime.  Moreover, we find that the
superpartners of the D0-brane, in this $r \ra 0$ limit, possess
neither spin nor D2-brane dipole moments.}


\begin{document} 

\section{Introduction}

At present, it is unclear as to what are the fundamental physical degrees of freedom
underlying M-theory.  Most of what we understand about M-theory
is based on the facts that it has eleven-dimensional supergravity as
its low energy limit and, via compactification on a circle, can be
related to type IIA string theory~\cite{Paul,Witten}.  While the
string of the type IIA theory was at one time regarded as the fundamental object,
it is now clear that this is not the basic degree of freedom
underpinning M-theory as a whole.  Of course, one may take the point of view that there is no \emph{truly}
fundamental physical degree of freedom but rather that, in different regions of the
M-theory moduli space, different degrees of freedom appear.

An interesting approach to M-theory, which may resolve at least some
of these conceptual issues, is Matrix theory: a non-relativistic supersymmetric quantum
mechanics of $N \times N$ matrix degrees of freedom, as considered in
an earlier guise in~\cite{CH} (see, e.g.,~\cite{Wati}
for a recent review).  Matrix theory was originally conjectured to be equivalent to
M-theory in the infinite momentum frame, in the limit that $N$ is
large~\cite{BFSS}.  The \emph{finite} $N$ version has been further
conjectured to be equivalent to the discrete light cone quantization
(DLCQ) of M-theory~\cite{Suss,Sei,Sen}.  The Hamiltonian of Matrix
theory is precisely the low-energy, or non-relativistic, limit of the
Hamiltonian describing a system of $N$ type IIA D0-branes (as such,
the corresponding action can be obtained via the null reduction of the
action describing the eleven-dimensional massless particle~\cite{Paul2}).  This makes
sense given that the D0-brane couples to the Ramond-Ramond (R-R) vector of the type IIA
theory, which is itself the Kaluza-Klein vector obtained
by dimensionally reducing M-theory on a circle.  In other words,
D0-branes are the ``partonic'', or fundamental, degrees of freedom
underlying Matrix theory.

A key fact about D0-branes is that they possess ``internal'' degrees
of freedom, which couple to spacetime fields.
More specifically, on the world-volume of a D0-brane there resides a simple
quantum mechanical theory which includes 16 fermionic operators
$\th$.  They generate an $SO(16)$ Clifford algebra, so may be
written as $2^{16/2} = 256$-dimensional gamma matrices~\cite{MTR}, i.e. a D0-brane has
256 internal degrees of freedom, or polarization states\footnote{D0-branes in different polarization states have different
interactions: in addition to the usual velocity-dependent terms, there
are also spin-dependent static forces, which were first studied
in~\cite{Harvey2,MSS}.  Other studies in the context of Matrix theory
include~\cite{PW,PSW1,PSW2}.}.  This space of states has
been constructed in~\cite{MTR}, and is just the space of
polarization states of the supergraviton in eleven dimensions.  In the
weak field approximation, the worldvolume fermions couple to small
fluctuations of the background metric, $h_{ab}$, NS-NS 2-form potential, $B_{ab}$, and R-R 1-form and 3-form
potentials, $C_a$ and $C_{abc}$, via the terms~\cite{MSS2,TR,MTR}
\be
\L_{D0} = -\frac{i}{8} \left( \del_i h_{tj} + \del_i C_j
\right) \bar{\th} \ga^{ij} \th + \frac{i}{16} \left( \del_i B_{jk} + \del_i C_{tjk}
\right) \bar{\th} \ga^{ijk} \th.
\label{eqn:action}
\ee

\noindent In other words, the internal degrees of freedom generate non-trivial long-range supergravity
fields.

\subsection{Bosonic p-branes and their superpartners}

The D0-brane solution of type IIA supergravity is an example of a more general class of extremal
$p$-brane.  These solitonic solutions of ten- and eleven-dimensional
supergravity have been much studied (see, e.g.,~\cite{Stelle,Marolf} for
reviews).  One is usually interested in purely bosonic solitons,
which nevertheless admit Killing spinors, such solutions thus being invariant under some
fraction of the 32 supersymmetries.  For a single $p$-brane, this
fraction is one half, which leaves 16 \emph{broken} supersymmetries.
They correspond to 16 zero mode fermions, the presence of which gives
rise to the entire BPS supermultiplet of $256$ states.  These
polarization states of the spinning $p$-brane fall into
representations of the little group of the respective brane so, as we
have already mentioned, the states of the spinning D0-brane match those of
the eleven-dimensional supergraviton; the little group in both
cases is $SO(9)$.

From the worldvolume perspective, the polarization state of a spinning $p$-brane is reflected in its
couplings to the background supergravity fields, via terms such as
(\ref{eqn:action}).  The generic state couples to \emph{all} such fields, not just the metric,
$(p+1)$-form potential and dilaton.  As can be seen from
(\ref{eqn:action}), certain states of the D0-brane, for
example, have dipole moments with respect to the R-R 3-form
potential~\cite{MTR}, in addition to moments with respect to the other
bosonic fields\footnote{It should be emphasized, however, that such
polarized branes are different in nature to the purely bosonic dielectric branes
of Myers~\cite{Myers}.  The dipole moments of the latter have their origin
in the non-abelian worldvolume theory of multiple D-branes, and the
resulting solutions are spatially extended.  The polarization
states considered herein, however, are states of a \emph{single}
brane.}.  We should further note that the various quadrupole couplings and
moments have also been worked out in~\cite{MSS2}.

Each such polarization state has a corresponding supergravity solution, which
displays the multipole moments in question.   These are the so-called ``superpartners''~\cite{AE}
of the purely bosonic solutions, and they can be generated by acting on the
latter with broken supersymmetry transformations, as
first discussed by Aichelburg and Embacher~\cite{AE}.  They exhibited the complete
supermultiplet containing the extreme Reissner-Nordstr\"{o}m multiple black
hole solutions in four-dimensional $\N = 2$ supergravity.  One considers the finite transformation
\be
{\bf \Phi} \ra {\bf \Phi'} = e^{\de_{\ep}} {\bf \Phi} = {\bf \Phi} +
\de_{\ep} {\bf \Phi} + \half \de_{\ep}^2 {\bf \Phi} + \ldots,
\label{eqn:transform}
\ee

\noindent where ${\bf \Phi}$ denotes the bosonic solution, and
$\de_{\ep}$ the action of a broken infinitesimal supersymmetry
transformation with parameter $\ep$.  Of course, a single such transformation leaves the field equations invariant at the
linearized level only, but the finite transformation (\ref{eqn:transform})
will be a symmetry of the full non-linear field equations.  

The first order variation, $\de_{\ep}$, generates fermionic ``hair'', and so a
non-vanishing supercharge.  Corrections to the bosonic fields, which
give rise to various dipole moments, are
generated by the second order variations, $\de_{\ep}^2$, 
and so on.  Since $\ep$ is a Grassman quantity, the series
(\ref{eqn:transform}) terminates at $\de_{\ep}^{16}$ for the ten- and
eleven-dimensional supergravities; $\ep$ has 16 independent
components in these cases.  Complete supermultiplets
containing the bosonic $p$-brane solutions of these theories are thus unlikely
to be found, but spin and magnetic dipole moments are readily
analysed; in addition to black hole and string states in four-dimensional $\N =4$ string
compactifications~\cite{DLR2}, superpartners of the
D0-brane~\cite{DLR,MTR}, the M2-brane~\cite{BKTW} and the
M5-brane~\cite{Win} have been studied using such techniques.

As localized solitons, $p$-branes possess bosonic zero modes
corresponding to broken translational symmetries, the associated collective
coordinates specifying the centre-of-mass
position of the brane in  question.  In a similar manner, the 16
broken supersymmetries correspond to fermionic zero modes so, in a sense, these fermionic moduli can be
thought of as fermionic ``collective coordinates''.  There is an
important difference, however~\cite{JR,Harvey,BKTW}: the fermionic zero modes satisfy non-trivial anti-commutation
relations, inherited from the anti-commutation relations of the
spacetime fermions.  Since they must be realised as operators, the
back-reaction of the fermionic zero modes on the supergravity fields results in operator-valued
expressions, and the background fields have a meaning
only in the sense of an expectation value for a given BPS
state~\cite{BKTW}.  Moreover, the possible polarization states form representations of the algebra
of fermionic zero mode operators, which is $SO(16)$ in the ten- and
eleven-dimensional examples.  Given this space of states, one can then
ask the question as to whether the original bosonic solution is \emph{exact}, i.e. whether
the back-reaction of the fermionic zero modes vanishes in the corresponding
BPS state, which will be a singlet under the
respective little group.  Since the $SO(16)$ vacuum state is an
$SO(8)$ singlet, the purely bosonic M2-brane is thus exact as a BPS state~\cite{BKTW}, although
this is not the case for the D0-brane: there is no $SO(9)$ singlet of
$SO(16)$, so the purely bosonic D0-brane soliton is not an exact BPS
state~\cite{DLR}.  Of course, the same can be said of the purely
bosonic eleven-dimensional supergraviton.

It is precisely the background supergravity fields in (\ref{eqn:action})
which are determined using these superpartner generating techniques.  With respect to the D0-brane, rather than working with
type IIA supergravity directly, an alternative as suggested in~\cite{MTR} is to work with the eleven-dimensional
gravitational, or pp-wave.  Our purpose here is to consider this
approach.  We compute the superpartners of the purely bosonic
pp-wave in the following section, and dimensionally reduce them in the standard
way~\cite{CW,HN} to give the superpartners of the D0-brane in section
3.  We find exact agreement with the results of ~\cite{DLR,MTR}.
Indeed, the fact that the gyromagnetic ratios of the D0-brane are both
equal to 1~\cite{DLR,MSS2,MTR} is a natural consequence of their
eleven-dimensional origins.  Such gyromagnetic ratios are
characteristic of Kaluza-Klein states~\cite{DLR}.  Given the
connections between the DLCQ of M-theory and Matrix theory at finite
$N$, it is of further interest to consider the dimensional reduction on a light-like
circle (or rather an asymptotically light-like circle).  This we do in section 4,
before concluding.  It would seem that physical properties of the
resulting ten-dimensional solution, which is not asymptotically flat,
are most easily analysed from this eleven-dimensional perspective.  Our conventions are given in an appendix.

\section{Eleven-dimensional supergravitons}

\subsection{The bosonic solution}

Eleven-dimensional supergravity~\cite{CJS} consists of the elfbein $E^{\ul{A}}_{~A}$, a 3-form
potential $A_{ABC}$, with 4-form field strength $F_{ABCD} = 4
\del_{[A} A_{BCD]}$, and a spin $3/2$ Majorana gravitino $\Psi_{A}$.
In our conventions, the Lagrangian and equations of motion are invariant under the
infinitesimal supersymmetry transformations
\ba
\de_{\ep} E^{\ul{A}}_{~A} &=& i \,\bar{\ep} \,\Ga^{\ul{A}} \Psi_{A}, \nonumber \\
\de_{\ep} A_{ABC} &=& -3 i \,\bar{\ep} \,\Ga_{[AB} \Psi_{C]},
\label{eqn:susy} \\
\de_{\ep} \Psi_A &=& \D_A (\tilde{\om}) \,\ep, \nonumber
\ea

\noindent where $\ep$ is an arbitrary anticommuting Majorana spinor
and
\ba
\D_A (\tilde{\om}) &=& \del_A + \quarter \tilde{\om}_{A\ul{A}\,\ul{B}}
\Ga^{\ul{A}\,\ul{B}} - \frac{1}{288} \left( \Ga_A^{~BCDE} - 8 \de_A^{B}
\Ga^{CDE} \right) \F_{BCDE}, \nonumber \\
\tilde{\om}_{A\ul{A}\,\ul{B}} &=& \om_{A\ul{A}\,\ul{B}}
-\frac{i}{2} \left( \bar{\Psi}_A \Ga_{\ul{B}} \Psi_{\ul{A}} -
\bar{\Psi}_A \Ga_{\ul{A}} \Psi_{\ul{B}} + \bar{\Psi}_{\ul{B}} \Ga_A
\Psi_{\ul{A}} \right), \\
\F_{ABCD} &=& F_{ABCD} - 3i \bar{\Psi}_{[A} \Ga_{BC} \Psi_{D]}. \nonumber
\ea

\noindent $\om_{A\ul{A}\,\ul{B}}$ denotes the standard spin
connection.  We also make use of the infinitesimal $SO(10,1)$ Lorentz
transformations which, in terms of the arbitrary infinitesimal
parameter $\Lambda_{\ul{A}\,\ul{B}} = -\Lambda_{\ul{B}\,\ul{A}}$, are
\ba
\de_L E^{\ul{A}}_{~A} &=& \Lambda^{\ul{A}}_{~\ul{B}} E^{\ul{B}}_{~A},
\nonumber \\
\de_L \Psi_A &=& \quarter \Lambda_{\ul{A}\,\ul{B}} \Ga^{\ul{A}\,\ul{B}}
\Psi_A. \\
\de_L A_{ABC} &=& 0. \nonumber
\ea

Switching off the 3-form and gravitino, we have pure
eleven-dimensional gravity, the equations of motion of
which admit solutions with a null Killing vector, as first discussed by Hull~\cite{Hull}.
They describe gravitational waves propagating at the speed of light.  Denoting the direction of propagation
by $z$, the metric can be written as
\be
ds^2 = -(2-H) dt^2 + H dz^2 + 2(1-H) dz dt + dx^i dx^i,
\label{eqn:wave}
\ee

\noindent where $x^i, i=1, \ldots, 9$ denote the Cartesian coordinates
on $\mathbb{R}^9$, the space transverse to the $tz$-plane.  The function $H$
is harmonic on this space, the asymptotically flat solution being given by
\be
H(r) = 1 + \frac{2}{7} \frac{\ka_{11}^2}{\Om_8} \frac{P}{r^7},
\label{eqn:H}
\ee

\noindent where $r = |x|$ is the radial coordinate on $\mathbb{R}^9$
and $\Om_8$ is the volume of a unit eight-sphere.  In general, the momentum density,
$P$, has a dependence on the retarded time, $x^- = t - z$, so
that the amplitude of the pp-wave varies across the wave-front.  The
ADM energy-momentum is then
\be
P^t = P^z = \int dz P(x^-).
\label{eqn:momentum}
\ee

\noindent As we are ultimately
interested in the dimensional reduction of the pp-wave in the $z$
direction, we assume $P = P^z/2\pi R = \mbox{const.}$, with
$R$ the compactification radius.  In this case, the solution describes
a plane-fronted parallel gravitational wave (pp-wave for short).  Making use of the null Killing
direction, one can also find pp-wave solutions with non-vanishing
3-form and gravitino~\cite{Hull}.  But as these are \emph{not} given by acting on the purely
bosonic solution (\ref{eqn:wave}) with supersymmetry
transformations~\cite{Hull}, they are not the type of solution we are
looking for.

Of course, the pp-wave (\ref{eqn:wave}) can be viewed as the extremal
limit of an infinitely boosted Schwarzschild
black hole, or rather of an infinitely boosted
uncharged black string~\cite{Gary,RT}.  To see this, take
\be
ds^2 = - \left( 1 - \frac{M}{r^7} \right)  dt'^2 + dz'^2 + \left( 1 -
\frac{M}{r^7} \right)\inv dr^2 + r^2 d \Om^2_8,
\ee

\noindent and perform the boost
\be
t' = \cosh \mu ~t - \sinh \mu ~z , \qquad z' = \cosh \mu ~z - \sinh \mu ~t.
\ee

\noindent One finds
\be
ds^2 = - \left( 1 - \frac{M}{r^7} \right) H\inv dt^2 + H \left( dz +
\coth \mu (H\inv - 1) dt \right)^2 + \left( 1 - \frac{M}{r^7}
\right)\inv dr^2 + r^2 d \Om^2_8,
\label{eqn:nonext_wave}
\ee

\noindent where
\be
H(r) = 1 + \sinh^2 \mu \frac{M}{r^7},
\ee

\noindent is the harmonic function associated with the wave.  It
carries momentum density $P \sim \sinh^2 \mu \,M$.  Keeping $P$ fixed whilst taking the
extremal limit, $M \ra 0$, requires an infinite boost, $\mu \ra
\infty$.  In this limit, the metric (\ref{eqn:nonext_wave}) takes the form (\ref{eqn:wave}) as
promised.

\subsection{Superpartners}

We analyse the pp-wave (\ref{eqn:wave}) in the pseudo-orthonormal basis:
\be
E^{\ul{t}} = H^{-1/2} dt, \qquad E^{\ul{z}} = H^{-1/2}(1-H) dt +
H^{1/2} dz, \qquad E^{\ul{i}} = dx^i,
\label{eqn:elfbein}
\ee

\noindent where
\be
ds^2 = - (E^{\ul{t}})^2 + (E^{\ul{z}})^2 + E^{\ul{i}} E^{\ul{i}},
\label{eqn:basis_metric}
\ee

\noindent and the triangular parameterization, $E^{\ul{t}}_{~z} = 0 =
E^{\ul{i}}_{~z}$, allows for the dimensional reduction in the $z$
direction.  The superpartners of the bosonic wave are generated by acting on the
solution (\ref{eqn:elfbein}) with the supersymmetry transformations
(\ref{eqn:susy}).  As explained in~\cite{CW,HN}, to restore the triangular form
of the elfbein, and to ensure the canonical form of the
ten-dimensional supersymmetry transformations, we must perform
compensating $SO(10,1)$ and $SO(9,1)$ Lorentz transformations.
Schematically~\cite{CW},
\be
\de_{\eta} (D=10) = \de_{\ep} (D=11) + \de_{L_1}
(\Lambda^{\ul{a}}_{~\ul{z}}) + \de_{L_2} (\Lambda^{\ul{a}}_{~\ul{b}}),
\ee

\noindent where the ten-dimensional supersymmetry parameter, $\eta$,
is related to $\ep$ as will become clear.  We thus consider the
overall transformation
\ba
\de_{\ep} E^{\ul{A}}_{~A} &=& ( \Lambda^{~\ul{A}}_{1~~\ul{B}} +
\Lambda^{~\ul{A}}_{2~~\ul{B}} ) E^{\ul{B}}_{~A} + i \,\bar{\ep}
\,\Ga^{\ul{A}} \Psi_{A}, \nonumber \\
\de_{\ep} \Psi_A &=& \quarter ( \Lambda_{1 \,\ul{A} \,\ul{B}} +
\Lambda_{2 \,\ul{A} \,\ul{B}} )  \Ga^{\ul{A}\,\ul{B}} + \D_A
(\tilde{\om}) \ep, \label{eqn:variation} \\ 
\de_{\ep} A_{ABC} &=& -3 i \,\bar{\ep} \,\Ga_{[AB} \Psi_{C]}, \nonumber
\ea

\noindent where~\cite{CW,HN}
\ba
\Lambda^{~\ul{a}}_{1~~\ul{z}} &=& -\,i \bar{\ep} \,\Ga^{\ul{a}}
\Psi_{\ul{z}}, \nonumber \\ 
\Lambda^{~\ul{a}}_{2~~\ul{b}} &=& \frac{i}{8} \,\bar{\ep}
\,\Ga^{\ul{a}}_{~\ul{b}} \,\Ga^{\ul{z}} \,\Psi_{\ul{z}}.
\ea

\noindent The resulting fields will then be given by (\ref{eqn:transform})
with $\de_{\ep}$ as in (\ref{eqn:variation}).

Since the original solution (\ref{eqn:wave}) has $\Psi_A = 0$, a single transformation generates fermions alone.
We have
\ba
\de_{\ep} \Psi_t &=& \del_t \ep - \half H^{-1/2} \del_i H
\,\Ga^{\ul{i}\,\ul{t}} P_+ \ep, \nonumber \\
\de_{\ep} \Psi_z &=& \del_z \ep + \half H^{-1/2} \del_i H
\,\Ga^{\ul{i}\,\ul{t}} P_+ \ep, \label{killing_eqns} \\
\de_{\ep} \Psi_i &=& \del_i \ep + \quarter H\inv \del_i H \ep - \half
H\inv \del_i H P_+ \ep, \nonumber
\ea

\noindent where the projection operators
\be
P_{\pm} = \half \left( 1 \pm \Ga^{\ul{t}} \,\Ga^{\ul{z}} \right),
\label{eqn:projector}
\ee

\noindent can be used to split the supersymmetry parameter as
$\ep = P_+ \ep + P_- \ep \equiv \ep_+ + \ep_-$.  The unbroken
supersymmetry parameters are given by $\ep =
H^{-1/4} \ep_-$, so that $P_+ \ep = 0$ and the bosonic solution
(\ref{eqn:wave}) does indeed preserve one half of the
supersymmetries.  It is then clear that any spinor $\ep = \E(x^i)
\ep_0$, for $\E \ne H^{-1/4}$ or $\ep_0 \ne \ep_-$ will generate a
gravitino.  However, if the associated supercharge is to be non-zero
and finite, we should choose $\ep_0 = \ep_+$, and demand
that $\E \ra 1$ as $r \ra \infty$~\cite{AE}.

The choice of function $\E$ is then a choice of gauge.  One way to fix
the gauge freedom is to impose the tracelessness condition, $\Ga^A
\de_{\ep} \Psi_A$ = 0, on the first order gravitino~\cite{BKO}, so that it
is a pure spin $3/2$ excitation.  On the other hand, perhaps a more
fundamental criterion is that the first order gravitino be
normalizable~\cite{Win}.  In many cases, these two restrictions
coincide~\cite{BKO}, but it is not clear whether they will in
general~\cite{Win} (see~\cite{Carlos} for a discussion of these
issues).  In the case at hand, we take $\E = H^{-1/4}$ in analogy with
the unbroken supersymmetries.  Substituting for $\ep = H^{-1/4} \ep_+$
in (\ref{killing_eqns}) gives
\ba
\de_{\ep} \Psi_t &=& -\half H^{-3/4} \del_i H \,\Ga^{\ul{i}\,\ul{t}}
\,\ep_+, \nonumber \\
\de_{\ep} \Psi_z &=& \half H^{-3/4} \del_i H \,\Ga^{\ul{i}\,\ul{t}}
\,\ep_+, \label{eqn:delta_psi} \\
\de_{\ep} \Psi_i &=& - \half H^{-5/4} \del_i H \,\ep_+. \nonumber
\ea

\noindent It is easy to check that this first
order gravitino is normalizable:
\be
\left| \de_{\ep} \Psi \right|^2 = \frac{1}{\ka_{11}^2} \int_{\S} d^{10} x \sqrt{g_{(10)}}
\Psi_A^{\dagger} \Psi_B g^{AB} = \half P^z \ep_+^{\dagger}
\ep_+,
\label{eqn:norm}
\ee

\noindent where $\S$ denotes a space-like hypersurface with induced
metric $g_{(10)}$.  This we take as justification for our choice of
$\E$ and $\ep_0$.  However, the tracelessness condition, $\Ga^A
\de_{\ep} \Psi_A$ = 0, is \emph{not} satisfied for this choice.

With these first order variations, a second application of the
transformations (\ref{eqn:variation}) generates further bosonic
fields, given in terms of fermion bilinears.  The second order
variation of the elfbein is
\ba
\de_{\ep}^2 E^{\ul{t}}_{~i} &=& \frac{i}{16} H^{-3/2} \del_j H \,\bar{\ep}_+
\,\Ga^{\ul{i}\,\ul{j}\,\ul{t}} \,\ep_+, \nonumber \\
\de_{\ep}^2 E^{\ul{i}}_{~t} &=& -\frac{7i}{16} H^{-2} \del_j H \,\bar{\ep}_+
\,\Ga^{\ul{i}\,\ul{j}\,\ul{t}} \,\ep_+, \label{eqn:delta2_E} \\
\de_{\ep}^2 E^{\ul{z}}_{~i} &=& \frac{i}{2} H^{-3/2} \del_j H \,\bar{\ep}_+
\,\Ga^{\ul{i}\,\ul{j}\,\ul{t}} \,\ep_+, \nonumber
\ea

\noindent which, up to $\O(\ep^4)$ terms, gives the metric
\be
ds^2 = -(2-H) dt^2 + H dz^2 + 2 (1-H) dz dt - \frac{i}{2} H\inv
\del_j H  \,\bar{\ep}_+ \,\Ga^{\ul{i}\,\ul{j}\,\ul{t}} \,\ep_+ ( dt - dz ) dx^i  + dx^i dx^i.
\label{eqn:delta_metric}
\ee

\noindent The fermion bilinears in the metric (\ref{eqn:delta_metric}), and in
the 3-form derived below, can be expressed in terms of $SO(9)$
creation and annihilation operators~\cite{MTR}; can one see explicitly
that these fields have a meaning only in the sense of an expectation
value for a given BPS state.

Since $\de_{\ep}^2 g_{ij} = 0$, and since $\de_{\ep}^2 g_{ti}$ falls
off too quickly to affect the relevant integrals at infinity, the
fields (\ref{eqn:delta2_E}) do not alter the ADM
energy-momentum and all members of the supermultiplet are massless.
However, the $g_{zi} = - g_{ti}$ off-diagonal components of the metric
give rise to a conserved angular momentum, $J_{\mu ij}$, where we denote the
``worldvolume'' directions by $\mu = t,z$.  As in~\cite{BKTW}, the
angular momentum carries a worldvolume index as well as a pair of
transverse indices indicating the plane of rotation.  It can easily be
read off from the metric: with~\cite{MP}
\be
g_{\mu i} \ra -\frac{\ka_{11}^2}{\Om_8} J_{\mu ij}
\frac{\hat{x}^j}{r^8},
\label{eqn:spin_defn}
\ee

\noindent as $r \ra \infty$, we have
\be
J_{\mu ij} = -\frac{i}{2} P \,\bar{\ep}_+ \,\Ga^{\ul{i}\,\ul{j}\,\ul{\mu}} \,\ep_+,
\label{eqn:spin}
\ee

\noindent which is an angular momentum density.  Note that $g_{zi} = -
g_{ti}$ gives $J_{zij} = - J_{tij}$.  Being generated
by a static fermion bilinear, the angular momentum is not of the
Kerr-type and is more rightly interpreted as an ``intrinsic'' angular
momentum or \emph{spin}~\cite{AE}.  Upon dimensional reduction in the $z$-direction,
$J_{tij}$ and $J_{zij}$ are respectively identified
as the spin and magnetic dipole moments of the D0-brane.

Turning to the 3-form potential, which does not change under the
Lorentz transformations, we find
\be
\de_{\ep}^2 A_{\mu ij} = \frac{i}{2}
H\inv \del_k H \,\bar{\ep}_+ \,\Ga^{\ul{i}\,\ul{j}\,\ul{k}\,\ul{\mu}}
\ep_+.
\label{eqn:delta2_A}
\ee

\noindent All states of the
supermultiplet thus have the BPS property $M=Q=0$, since none are charged with
respect to the 3-form: the fields (\ref{eqn:delta2_A}) die off too quickly to
affect integrals of the field strength at infinity.  There is, however, a
non-vanishing dipole moment associated with the 3-form potential (\ref{eqn:delta2_A}).
This is a 4-index tensor, $\mu_{\mu ijk}$, with a single
worldvolume and three transverse indices.  With~\cite{BKTW}
\be
A_{\mu ij} \ra \frac{\ka_{11}^2}{\Om_8} \mu_{\mu ijk}
\frac{\hat{x}^k}{r^8},
\label{eqn:dipole_defn}
\ee

\noindent as $r \ra \infty$, we have
\be
\mu_{ \mu ijk} = - \frac{i}{2} P \,\ep_+
\Ga^{\ul{i}\,\ul{j}\,\ul{k}\,\ul{\mu}} \,\ep_+,
\label{eqn:dipole}
\ee

\noindent so that $\mu_{tijk} = - \mu_{zijk}$, as for the angular
momentum (\ref{eqn:spin}).  In ten dimensions, the dipole moments $\mu_{tijk}$ and
$\mu_{zijk}$ are respectively identified as an electric dipole moment with respect
to the R-R 3-form potential, and a magnetic dipole moment with respect to the NS-NS $B$ field.

The superpartners also carry a supercharge density, given by the
surface integral~\cite{Tei}
\be
Q = \frac{i}{2\ka_{11}^2} \oint_{\infty} dS_i \,\Ga^{i\al} \,\Psi_{\al},
\ee

\noindent where $\al = i,z$ runs over all \emph{spatial} directions,
and the gravitino is given by (\ref{eqn:delta_psi}).  We find
\be
Q = -\frac{i}{2} P \,\ep_+,
\ee

\noindent the form of which should be expected~\cite{AE}.
Higher-order corrections to the gravitino will not alter this value of
$Q$.

\section{Polarized D0-branes}

To dimensionally reduce in the $z$-direction, we make the usual
Kaluza-Klein ansatz for the elfbein:
\be
E^{\ul{A}}_{~A} = \left( \begin{array}{cc} E^{\ul{a}}_{~a} &
E^{\ul{z}}_{~a} \\ 0 & E^{\ul{z}}_{~z} \end{array} \right) = 
\left( \begin{array}{cc} e^{-\p/12} e^{\ul{a}}_{~a} & e^{2\p/3} C_a
\\ 0 & e^{2\p/3} \end{array} \right),
\label{eqn:KK_elfbein}
\ee

\noindent with inverse
\be
E^A_{~\ul{A}} = \left( \begin{array}{cc} E^a_{~\ul{a}} & E^z_{~\ul{a}} \\ 0 & E^z_{~\ul{z}} \end{array} \right) = 
\left( \begin{array}{cc} e^{\p/12} e^a_{~\ul{a}} & -e^{\p/12} C_{\ul{a}}
\\ 0 & e^{-2\p/3} \end{array} \right).
\ee

\noindent The 3-form potential reduces to the ten-dimensional R-R
3-form potential, $C_{abc}$, and NS-NS 2-form potential $B_{ab}$:
\be
A_{ABC} = \left( A_{abc}, A_{abz} \right) = \left( C_{abc}, B_{ab}
\right).
\label{eqn:KK_3form}
\ee

\noindent It is important to note that the above fields are those of
the full superpartner solutions.  Take, for example, the $E^{\ul{z}}_{~z} = e^{2\p/3}$ component
of the elfbein.  This is really the transformed field
\be
E'^{\ul{z}}_{~z} =  E^{\ul{z}}_{~z} \left( 1 + E^z_{~\ul{z}} \de_{\ep}
E^{\ul{z}}_{~z} +  E^z_{~\ul{z}} \half \de_{\ep}^2 E^{\ul{z}}_{~z} + \O(\ep^3) \right),
\ee

\noindent which should be equated with
\be
e^{2\p'/3} = e^{2\p/3} \left( 1 + \frac{2}{3} \de_{\ep} \p +
\frac{4}{9} \half \de_{\ep}^2 \p + \O(\ep^3) \right),
\ee

\noindent so that $e^{2\p} = H^{3/2}$ as required.  However, since
the $E^{\ul{z}}_{~z}$ component of the elfbein receives no
corrections, neither does the dilaton; and this simplifies matters
considerably.  As expected, the original pp-wave (\ref{eqn:wave}) gives the purely bosonic D0-brane solution
\ba
ds^2 &=& -H^{-7/8} dt^2 + H^{1/8} dx^i dx^i, \nonumber \\
C_{t} &=& H\inv - 1, \label{eqn:D0} \\
e^{2\p} &=& H^{3/2}. \nonumber
\ea

\noindent The longitudinal momentum is quantized and equal to the mass, $T_0$, of the D0-branes: $P^z = N/R =
N/(g_s \sqrt{\al'}) = T_0$.  Then the harmonic function $H$ is
\be
H(r) = 1 + 60 \pi^3 g_s \frac{ \al'^{7/2} N}{r^7}.
\ee

The first order variations (\ref{eqn:delta_psi}) of $\Psi_A$ reduce to
corresponding variations of the ten-dimensional fermions.  The ten-dimensional Dirac matrices are given
by
\be
\Ga^{\ul{A}} = \left( \Ga^{\ul{a}}, \Ga^{\ul{z}} \right) = \left(
\ga^{\ul{a}}, \ga^{11} \right),
\label{eqn:dirac}
\ee

\noindent where $\ga^{11} = \ga^{\ul{t}} \ldots \ga^{\ul{9}}$ is the
ten-dimensional chirality operator, and the gravitino reduces in a
similar manner:
\be
\Psi_{\ul{A}} = \left( \Psi_{\ul{a}}, \Psi_{\ul{z}} \right).
\ee

\noindent The ten-dimensional gravitino, $\psi_a = e^{\ul{a}}_{~a}
\psi_{\ul{a}}$, and dilatino, $\la$, are then given by~\cite{CW,HN}
\ba
\Psi_{\ul{a}} &=& e^{\p/24} \left( \psi_{\ul{a}} - \frac{\sqrt{2}}{12}
\ga_{\ul{a}} \ga^{11} \la \right), \\
\Psi_{\ul{z}} &=& \frac{2\sqrt{2}}{3} e^{\p/24} \la,
\ea

\noindent and the ten-dimensional supersymmetry parameter is $\eta =
e^{\p/24} \ep$~\cite{CW,HN}.  This latter implies $\eta = H^{-7/32}
\ep_+$ where, in ten-dimensional language
\be
P_+ \ep_+ = \half \left( 1 + \ga^{\ul{t}} \,\ga^{11} \right) \ep_+ =
\ep_+,
\label{eqn:projectors_ten}
\ee

\noindent as should be expected~\cite{DLR,MTR}.  The variations of the
ten-dimensional fermions are thus
\ba
\de_{\eta} \la &=& \frac{3\sqrt{2}}{8} H^{-41/32} \del_i H
\,\ga^{\ul{i}\,\ul{t}} \,\ep_+, \nonumber \\
\de_{\eta} \psi_t &=& -\frac{7}{16} H^{-55/32} \del_i H
\,\ga^{\ul{i}\,\ul{t}} \,\ep_+, \label{eqn:delta_fermions} \\
\de_{\eta} \psi_i &=& \frac{1}{16} H^{-39/32} \del_j H (\ga^{\ul{i}\,\ul{j}}
- 7 \de^{ij} ) \ep_+, \nonumber
\ea

\noindent which agree precisely with the results of~\cite{DLR,MTR}.
Both the dilatino and gravitino are normalizable but, as in the
eleven-dimensional case, the gravitino does not obey the condition
$\ga^a \de_{\eta} \psi_a = 0$.

The second order corrections (\ref{eqn:delta2_E}) and
(\ref{eqn:delta2_A}) of the elfbein and 3-form reduce respectively to
second order variations of the zehnbein and R-R vector, and the R-R
3-form and NS-NS $B$ field.  We find
\ba
\de_{\eta}^2 e^{\ul{t}}_{~i} &=& \frac{i}{16} H^{-23/16} \del_j H \,\bar{\ep}_+
\,\ga^{\ul{i}\,\ul{j}\,\ul{t}} \,\ep_+, \nonumber \\
\de_{\eta}^2 e^{\ul{i}}_{~t} &=& -\frac{7i}{16} H^{-31/16} \del_j H \,\bar{\ep}_+
\,\ga^{\ul{i}\,\ul{j}\,\ul{t}} \,\ep_+, \nonumber \\
\de_{\eta}^2 C_i &=& \frac{i}{2} H^{-2} \del_j H \,\bar{\ep}_+
\,\ga^{\ul{i}\,\ul{j}\,\ul{t}} \,\ep_+, \label{eqn:delta_bosons} \\
\de_{\eta}^2 B_{ij} &=& -\frac{i}{2} H\inv \del_k H \,\bar{\ep}_+
\,\ga^{\ul{i}\,\ul{j}\,\ul{k}\,\ul{t}} \,\ep_+, \nonumber \\
\de_{\eta}^2 C_{tij} &=& \frac{i}{2} H\inv \del_k H \,\bar{\ep}_+
\,\ga^{\ul{i}\,\ul{j}\,\ul{k}\,\ul{t}} \,\ep_+. \nonumber
\ea

\noindent These fields again agree precisely with the results
of~\cite{DLR,MTR}.  The cross term $g_{ti}$ in the resulting D0-brane
metric generates an angular momentum, $J_{ij}$, in the same manner as
above, and the 1-form $C_i$ generates a magnetic dipole moment,
$\mu_{ij}$.  They are just the dimensional reduction of the spin
$J_{\mu ij}$, as in (\ref{eqn:spin}).  With
\be
J_{tij} = \frac{1}{2\pi R}
J_{ij}, \qquad J_{zij} = \frac{1}{2\pi R} \mu_{ij},
\ee

\noindent we have
\be
J_{ij} = - \mu_{ij} = - \frac{i}{2} T_0 \,\bar{\ep}_+
\,\ga^{\ul{i}\,\ul{j}\,\ul{t}} \,\ep_+.
\ee

\noindent The gyromagnetic ratio, $g$, of these two moments is given in
general via~\cite{DLR}
\be
\mu_{ij} = g \frac{Q}{2M} J_{ij},
\ee

\noindent for a particle of mass $M=T_0$ and charge $Q$.  Taking care
to account for the implicit factor of $\ka_{10}^2/\Om_8$ in our definition of $\mu_{ij}$, relative
to~\cite{DLR}, we have that $Q=-2T_0$.  In other words,
$g=1$~\cite{DLR}.  This is thus a natural consequence of the
M-theoretic origin of the polarized D0-brane: both the off-diagonal
terms in the metric, and the R-R magnetic potential have a common
origin in eleven dimensions.

Finally, there is a magnetic dipole moment, $\mu_{ijk}$, associated with $B_{ij}$
and an electric dipole moment, $d_{ijk}$, associated with $C_{tij}$.
Both come from the eleven-dimensional dipole moments
(\ref{eqn:dipole}).  We have
\be
\mu_{tijk} = \frac{1}{2\pi R} d_{ijk}, \qquad \mu_{zij} = \frac{1}{2\pi R} \mu_{ijk},
\ee

\noindent so that
\be
d_{ijk} = - \mu_{ijk} = - \frac{i}{2} T_0 \,\bar{\ep}_+ \,\ga^{\ul{i}\,\ul{j}\,\ul{k}\,\ul{t}} \,\ep_+.
\ee

\noindent The gyromagnetic ratio of these two ten-dimensional
dipole moments is again $g=1$~\cite{MTR}, and this is again due to the fact
that both moments have a common eleven-dimensional origin.

\section{On the asymptotic light cone}

\subsection{A tale of two bases}

To discuss the light-like compactification of the bosonic pp-wave and
its superpartners, we define the coordinates
\be
x^+ = -\half (t + z), \qquad x^- = t - z,
\ee

\noindent so that the pp-wave metric (\ref{eqn:wave}) becomes
\be
ds^2 = 2 dx^+ dx^- + F dx^{-2} + dx^i dx^i,
\label{eqn:wave_pm}
\ee

\noindent where
\be
F(r) = H(r) - 1 =  \frac{2}{7} \frac{\ka_{11}^2}{\Om_8} \frac{P}{r^7},
\ee

\noindent ensures asymptotic flatness.  Since
$F \ra 0$ as $r \ra \infty$, $x^+$ and $x^-$ are light cone
coordinates only at infinity in the transverse space.  In particular,
our time coordinate $x^+$ is related to the light cone time $x_{LC}^+$ via
\be
x^+ = x^+_{LC} - \frac{F}{2} x^-,
\ee

\noindent and $x^-$ is space-like everywhere except as $r \ra \infty$.  In terms of the
energy-momentum tensor, $T^{AB}$, the ADM energy-momentum is
\be
P^{A} = \int dx^- d^9 x \,T^{+ A},
\ee

\noindent which gives the light cone energy $P^- = 0$.  The
longitudinal light cone momentum is
\be
P^+ = \int dx^- P(x^-),
\ee

\noindent and we will again assume that the momentum
density, $P(x^-) = \mbox{const.}$

Perhaps the most natural basis to use in the analysis of the spacetime
(\ref{eqn:wave_pm}) is\footnote{We drop the underlines on the $\pm$
tangent space indices in the following.  To avoid confusion, note that
gamma matrices are written in terms of tangent space components, and the gravitino in
terms of curved space components.}
\be
E^+ = dx^+ + \half F dx^-, \qquad E^- = dx^-, \qquad E^{\ul{i}} =
dx^i,
\label{eqn:basis_pm}
\ee

\noindent so that
\be
ds^2 = 2 E^+ E^- + E^{\ul{i}} E^{\ul{i}}.
\ee

\noindent Then the Killing spinors are given by $\ep = \ep_-$, where $\Ga^- \ep_- =
0$.  With the broken supersymmetries $\ep = \ep_+$, where $\Ga^+ \ep_+ =
0$, there is a single
non-zero component of the gravitino:
\be
\de_{\ep} \Psi_- = - \quarter \del_i F \,\Ga^{\ul{i}} \Ga^- \,\ep_+.
\ee

\noindent This is manifestly normalizable -- it is null -- and
satisfies the tracelessness condition $\Ga^A \de_{\ep} \Psi_A = 0$, by
virtue of the fact that $(\Ga^-)^2 = 0$.  It is interesting to see
that these two properties coincide here, when they did not above (nor
will they below).  A second order supersymmetry transformation, with
\emph{no} compensating Lorentz transformations, gives just two
corrections to the bosonic fields:
\ba
\de_{\ep}^2 E^{\ul{i}}_{~-} &=& -\frac{i}{4} \del_j F \,\bar{\ep}_+
\Ga^{\ul{i}\,\ul{j}-} \ep_+, \nonumber \\
\de_{\ep}^2 A_{ij-} &=& \frac{i}{4} \del_k F \,\bar{\ep}_+
\Ga^{\ul{i}\,\ul{j}\,\ul{k}-} \ep_+.
\ea

\noindent The 3-form and metric thus have the same asymptotic
structure as in section 2.

Despite the simplicity of working in this basis, it is ill-suited to
dimensional reduction along $x^-$.  To perform such a compactification,
we must rather work with the basis
\be
E^{\ul{t}} = F^{-1/2} dx^+, \qquad E^{\ul{z}} = F^{-1/2} dx^+ +
F^{1/2} dx^-, \qquad E^{\ul{i}} = dx^i,
\label{eqn:basis_diag}
\ee

\noindent in terms of which the metric (\ref{eqn:wave_pm}) is again given by
(\ref{eqn:basis_metric}).  Such a choice has its own problems, however,
the root cause of which seems to be the following.  For the basis
(\ref{eqn:basis_pm}), we have $E^+ \ra dx^+$ and $E^- \ra dx^-$ as $r \ra \infty$, so the metric on the
$x^{\pm}$ plane is $ds^2 \ra 2dx^+ dx^-$ as expected.  On the other
hand, for the basis (\ref{eqn:basis_diag}) we na\"{\i}vely have
$E^{\ul{t}} \ra F^{-1/2} dx^+$ and $E^{\ul{z}} \ra F^{-1/2} dx^+$,
so that $ds^2 \ra 0$.  The metric is null at infinity.  We will
attempt to make sense of this below but, in the meantime, let us
proceed.

Working with the basis (\ref{eqn:basis_diag}), the Killing spinors are $\ep =
F^{-1/4} \ep_-$.  The natural broken supersymmetry parameters are then $\ep =
F^{-1/4} \ep_+$ where $P_+ \ep_+ = \ep_+$, with $P_+$ as in
(\ref{eqn:projector}).  This choice gives
\ba
\de_{\ep} \Psi_- &=& \half F^{-3/4} \del_i F \,\Ga^{\ul{i}\,\ul{t}}
\,\ep_+, \nonumber \\
\de_{\ep} \Psi_i &=& - \half F^{-5/4} \del_i F \,\ep_+.
\label{eqn:delta_psi_pm}
\ea

\noindent This gravitino is now \emph{non}-normalizable, however, this being due to the fact that the
zero mode parameter $\ep = F^{-1/4} \ep_+$ blows up at infinity,
rather than approaching a constant.  The choice of multiplicative
factor does not satisfy the condition discussed above \emph{viz}, that
it approach $1$ as $r \ra \infty$.  The second order variations
\ba
\de_{\ep}^2 E^{\ul{t}}_{~i} &=& \frac{i}{16} F^{-3/2} \del_j F \,\bar{\ep}_+
\,\Ga^{\ul{i}\,\ul{j}\,\ul{t}} \,\ep_+, \nonumber \\
\de_{\ep}^2 E^{\ul{i}}_{~+} &=& -\frac{7i}{16} F^{-2} \del_j F \,\bar{\ep}_+
\,\Ga^{\ul{i}\,\ul{j}\,\ul{t}} \,\ep_+, \nonumber \\
\de_{\ep}^2 E^{\ul{z}}_{~i} &=& \frac{i}{2} F^{-3/2} \del_j F \,\bar{\ep}_+
\,\Ga^{\ul{i}\,\ul{j}\,\ul{t}} \,\ep_+, \label{eqn:delta_bosons_pm} \\
\de_{\ep}^2 A_{ij-} &=& \frac{i}{2} F\inv \del_k F \,\bar{\ep}_+ \,\Ga^{\ul{i}\,\ul{j}\,\ul{k}\,\ul{t}}
\ep_+, \nonumber
\ea

\noindent give the modified metric
\be
ds^2 = 2 dx^+ dx^- + F dx^{-2} + dx^i dx^i + \frac{i}{2} F\inv \del_j
F \,\bar{\ep}_+ \,\Ga^{\ul{i}\,\ul{j}\,\ul{t}} \,\ep_+ dx^- dx^i,
\ee

\noindent in which the problem manifests itself again.  That is, the
angular momentum and dipole moments as given by (\ref{eqn:spin_defn})
and (\ref{eqn:dipole_defn}) respectively, are no longer defined; we have $F\inv \ra r^7/k$ as $r \ra \infty$,
instead of $H\inv \ra 1$, so the corrections to the bosonic fields do
not fall off fast enough at infinity.  On the face of it, we should simply
choose a different gauge for the fermionic zero modes.  However, it
would seem that these problems are due rather to
our choice of basis.  After all, the basis (\ref{eqn:basis_diag})
na\"{\i}vely gives a null metric at infinity, rather than a flat one
(in light cone coordinates).  This, in turn, is related to the fact that
compactification of $x^-$ gives the $r \ra 0$ ``near-horizon'' limit of
the D0-brane solution, which is no longer asymptotically flat.
Quantities such as angular momentum and dipole moments are potentially ill-defined in
such spacetimes.

\subsection{Dimensional reduction on the asymptotic light cone}

We are interested here in the dimensional reduction along $x^-$ of the
pp-wave and its superpartners of the previous subsection.  Such a
light-like compactification of M-theory is related to finite $N$
Matrix theory~\cite{Suss,Sei,Sen}.  Now in the standard
reduction described in section 3, we identified $z \sim z + 2\pi R_s$, but the proper
circumference, $L_s$ of the space-like circle is
\be
L_s = H^{1/2} 2 \pi R_s,
\ee

\noindent so that $L_s (r \ra \infty) = 2 \pi R_s$.  However, if we
identify $x^- \sim x^- + 2\pi R_l$, the
proper circumference, $L_l$, of the light-like circle is
\be
L_l = F^{1/2} 2 \pi R_l,
\ee

\noindent and so $L_l (r \ra \infty) = 0$.  The compact
direction is thus only \emph{asymptotically} light-like.  That our
basis (\ref{eqn:basis_diag}) gives a null metric at infinity should
perhaps then be expected.

The two compactifications are related~\cite{Sei}: in the limit $R_s
\ra 0$, the space-like compactification is related to
the light-like one by an infinite boost along $z$, with velocity
\be
v = \frac{R_l}{\sqrt{R_l^2 + 4R_s^2}} \approx 1 -
2\frac{R_s^2}{R_l^2}.
\label{eqn:boost}
\ee

\noindent Indeed, the two metrics (\ref{eqn:wave}) and
(\ref{eqn:wave_pm}) are related by just such a boost, provided that we
replace the original momentum density $P = 1/(2\pi R_s) N/R_s$ with
$P = 1/(2\pi R_l) N/R_l$.  More specifically, the light-like
compactification of M-theory, with gravitons carrying momentum $P^+ =
N/R_l$ around the light-like circle, is related by the
boost (\ref{eqn:boost}) to M-theory on the space-like circle, in the limits $R_s \ra
0$, $\al' \ra 0$ and  $U = R_s/\al' = \mbox{fixed}$~\cite{Sei,Sen}.  This is just the Maldacena
decoupling limit~\cite{Mal} applied to D0-branes, as considered
in~\cite{IMSY,BST}.  Supergravity in this background is then dual to
Matrix theory at finite $N$.

There are some subtleties, however.  Firstly, as we have already seen, the compactification
is only asymptotically light-like~\cite{BGL,Pol}.  More importantly, one should
be careful about the regimes of validity of the different
descriptions, the key point being that the radius of the eleventh dimension is set by the
dilaton~\cite{BGL}.  The story is explained in detail in~\cite{Pol}:
for large distances $r > l_p N^{1/3}$ (the UV of the gauge theory), the effective coupling constant of
the gauge theory is small, so that Matrix perturbation theory is
valid.  As $r$ decreases (moving toward the IR), the Matrix theory becomes strongly
coupled.  For $l_p N^{1/7} < r < l_p N^{1/3}$, supergravity in the
$r\ra 0$ limit of the D0-brane solution is the better description,
since the spacetime curvature
is small compared to the string scale, and the string coupling $e^{\phi}
<< 1$.  For smaller
distances still, this solution becomes eleven-dimensional, since the
dilaton is blowing up with
the effective gauge theory coupling.  The relevant description, for
$l_p N^{1/9} < r < l_p N^{1/7}$, is that
of a pp-wave on an asymptotically light-like circle, which is what we
consider here.  This description in turn breaks down at $r = l_p N^{1/9}$,
at which point the radius of the eleventh dimension is itself equal to
$r$, and the boosted black string of section 2 becomes unstable~\cite{IMSY}.

To compactify along a light-like circle, one considers the limit of a compactification on a space-like circle,
which is almost light-like~\cite{Sei,HP}.  In other words, one ``regularizes''
the original metric such that the coordinate in question is
everywhere space-like, performs the standard compactification of the
space-like coordinate, and only then does one take the light-like
limit~\cite{HP}.  In the case at hand, the function $F$ acts as just such a regulator~\cite{HK},
except as $r \ra \infty$.  To ensure that $x^-$ is \emph{everywhere} space-like, we thus write
\be
F = \frac{R_s}{R_l} + \frac{k}{r^7},
\label{eqn:F_reg}
\ee

\noindent perform the compactification in the standard way, and then
take the limit $R_s \ra 0$.  This procedure is thus essentially equivalent to the
compactification along $z$ as above, but with the $1$ finally dropped from
the harmonic function $H$~\cite{BBPT,Kraus,HKS}.  Indeed, it has been
shown explicitly in~\cite{BBPT,Paul2} that compactification along $x^-$ of the pp-wave
(\ref{eqn:wave_pm}) does indeed give the short distance, or
``near-horizon'', limit of the D0-brane solution (\ref{eqn:D0}).
Moreover, the null reduction of the action describing a massless
particle in eleven dimensions gives that describing a non-relativistic
D0-brane~\cite{Paul2}, which is precisely the Matrix theory action, up
to the $U(N)$ gauge symmetry.  We should also note that the strange
behaviour of the basis (\ref{eqn:basis_diag}) is cured via this
regularization of the function $F$.  With $F$ as in (\ref{eqn:F_reg}),
we have $E^{\ul{t}} \ra \sqrt{R_l/R_s} dx^+$ and $E^{\ul{z}} \ra
\sqrt{R_l/R_s} dx^+ + \sqrt{R_s/R_l} dx^-$, so that $ds^2 \ra 2dx^+
dx^-$ as required.

Now we can see why the gravitino of the previous subsection is not
normalizable, and why the
problems in defining the spin and dipole moments occur.  With $F$ as in (\ref{eqn:F_reg}), the norm of the
gravitino (\ref{eqn:delta_psi_pm}) is
\be
\left| \de_{\ep} \Psi \right|^2 = \half \frac{R_l}{R_s} P^+
\ep^{\dagger}_+ \ep_+,
\ee

\noindent where $P^+ = N/R_l$.  The infinite norm is then due to
taking the light-like limit $R_s \ra 0$.  But this is precisely what
we would find in the space-like compactification on a circle of
vanishing radius: with $P^z = N/R_s$, the norm of the gravitino in
(\ref{eqn:norm}) is also infinite in the limit $R_s \ra 0$.  As the latter is
Lorentz-equivalent to the former, the fact that the gravitino
(\ref{eqn:delta_psi_pm}) of the previous subsection is
non-normalizable should not come as any surprise!

Similar comments can be made concerning both the spin, $J_{-ij}$, and dipole
moment, $\mu_{-ijk}$ generated by the second order variations
(\ref{eqn:delta_bosons_pm}).  Since we now have $F \ra R_s/R_l$ as $r
\ra 0$, these physical quantities are well-defined, since the
corrections to the bosonic fields then fall off as $1/r^8$ as
required.  However, both quantities are then formally infinite in the $R_s
\ra 0$ limit, since they go like $1/R_s$:
\ba
J_{-ij} &=& \frac{i}{2} P^+ \frac{R_l}{R_s} \,\bar{\ep}_+
\,\Ga^{\ul{i}\,\ul{j}\,\ul{t}} \,\ep_+, \nonumber \\
\mu_{-ijk} &=& \frac{i}{2} P^+ \frac{R_l}{R_s} \,\bar{\ep}_+ \,\Ga^{\ul{i}\,\ul{j}\,\ul{k}\,\ul{t}}
\ep_+.
\label{eqn:moments_pm}
\ea

\noindent But this is just what we would expect upon taking the $R_s \ra 0$ limit of the
corresponding quantities in subsection 2.2.

\subsection{D0-branes as $r \ra 0$, and their superpartners}
 
Dimensional reduction along $x^-$ is now straightforward.  With
the Kaluza-Klein ans\"{a}tze as in (\ref{eqn:KK_elfbein}) and
(\ref{eqn:KK_3form}), but with $z$ replaced by $-$, the
purely bosonic solution (\ref{eqn:wave_pm}) reduces to
\ba
ds^2 &=& - F^{-7/8} dx^{+2} + F^{1/8} dx^i dx^i, \nonumber \\
C_{+} &=& F\inv, \label{eqn:D0_near} \\
e^{2\p} &=& F^{3/2}, \nonumber
\ea

\noindent where now $F = k/r^7$.  This is precisely the decoupling limit of the D0-brane
solution (\ref{eqn:D0}), as in~\cite{BBPT,Paul2}, and a conformal
transformation to the ``dual'' frame takes the metric to that of
AdS$_2$ $\times S^8$~\cite{BST}.  One might be concerned
that the $r \ra 0$ limit of the R-R 1-form potential in (\ref{eqn:D0})
should give a constant, and not $C_{+} = F\inv$.  However, one should instead
consider the corresponding limit of the field strength $F_{rt} =
\del_r(H\inv) \ra F_{r+} = 7r^6/k$, for which the 1-form above \emph{is} the
corresponding potential.  As to the ten-dimensional superpartners,
with the reduction formulae (\ref{eqn:dirac})--(\ref{eqn:projectors_ten}) we
find the fermions
\ba
\de_{\eta} \la &=& \frac{3\sqrt{2}}{8} F^{-41/32} \del_i F
\,\ga^{\ul{i}\,\ul{t}} \,\ep_+, \nonumber \\
\de_{\eta} \psi_+ &=& -\frac{7}{16} F^{-55/32} \del_i H
\,\ga^{\ul{i}\,\ul{t}} \,\ep_+, \\
\de_{\eta} \psi_i &=& \frac{1}{16} F^{-39/32} \del_j F (\ga^{\ul{i}\,\ul{j}}
- 7 \de^{ij} ) \ep_+, \nonumber
\ea

\noindent In other words, the first order supersymmetry variations of
the solution (\ref{eqn:D0_near}) are precisely the $r \ra 0$ limits
of the first order variations, (\ref{eqn:delta_fermions}), of the
D0-brane.  At this order, taking the $r \ra 0$ limit of the
D0-brane induces a corresponding limit in the entire supermultiplet.

This makes sense from the ten-dimensional perspective.  Replacing $H$
with $F$ in the purely bosonic D0-brane solution just gives a
non-asymptotically flat solution of type IIA supergravity.  It
still preserves $1/2$ of the supersymmetries, the Killing spinors now being
$\eta = F^{-7/32} \ep_-$.  Taking the broken supersymmetry parameters to be $\eta
= F^{-7/32} \ep_+$, we do indeed find the above fermions.  One might
object that this choice is invalid since it does not approach a
constant at infinity.  But then the solution is not asymptotically
flat, so quantities such as the supercharge are ill-defined.

Up to exchanging $H$ for $F$, the second order variations are unchanged relative to (\ref{eqn:delta_bosons}):
\ba
\de_{\eta}^2 e^{\ul{t}}_{~i} &=& \frac{i}{16} F^{-23/16} \del_j F \,\bar{\ep}_+
\,\ga^{\ul{i}\,\ul{j}\,\ul{t}} \,\ep_+, \nonumber \\
\de_{\eta}^2 e^{\ul{i}}_{~+} &=& -\frac{7i}{16} F^{-31/16} \del_j F \,\bar{\ep}_+
\,\ga^{\ul{i}\,\ul{j}\,\ul{t}} \,\ep_+, \nonumber \\
\de_{\eta}^2 C_i &=& \frac{i}{2} F^{-2} \del_j F \,\bar{\ep}_+
\,\ga^{\ul{i}\,\ul{j}\,\ul{t}} \,\ep_+, \\
\de_{\eta}^2 B_{ij} &=& -\frac{i}{2} F\inv \del_k F \,\bar{\ep}_+
\,\ga^{\ul{i}\,\ul{j}\,\ul{k}\,\ul{t}} \,\ep_+, \nonumber
\ea

\noindent although there is no electric R-R 3-form
potential.  It is straightforward to compute these corrections to the
solution (\ref{eqn:D0_near}) within type IIA supergravity directly,
and we do indeed find that $\de_{\eta}^2 C_{+ij} = 0$.  This is due
to the fact that $C_+ = F\inv \ne F\inv - 1$ in the purely bosonic
solution.  In the D0-brane solution, the $-1$ is a pure gauge
term which ensures that the potential vanishes at infinity.  In this
case, since the spacetime is not asymptotically flat, there is no such
requirement on the behaviour of $C_+$.  Of course, the absence of the
R-R 3-form is correlated with the fact that there is no electric
dipole moment for this field: upon dimensional reduction, $\mu_{-ijk}$
in (\ref{eqn:moments_pm})
gives a magnetic dipole moment associated with $B_{ij}$ only.  In a
similar manner, $J_{-ij}$ gives a magnetic dipole moment associated
with $C_i$, but no ten-dimensional spin (despite the fact that there
is a corresponding cross-term in the metric).  In other words, the
near-horizon solution of the D0-brane has no spin, and no D2-brane
dipole moment.  It has magnetic dipole moments alone.

\section{Discussion}

We have computed the superpartners of the purely bosonic
eleven-dimensional pp-wave in two coordinate systems, one adapted to a
space-like dimensional reduction along $z$, the other adapted to a
light-like dimensional reduction along $x^-$.  The former gives the
bosonic D0-brane and its superpartners, the latter gives the $r \ra 0$
limit of these solutions.  This ``near-horizon'' limit of the D0-brane
is a non-asymptotically flat spacetime, so its physical ADM-like
properties are ill-defined.  This is reflected in the fact that the
eleven-dimensional basis suited to the light-like dimensional
reduction must be regularized in order for the compactification to be
carried out, and so that the dipole moments of the supergraviton
spacetimes can be defined.  Even then, the first order gravitino as
well as these dipole moments are formally infinite, but we have argued that
this is to be expected.  The light-like compactification is
Lorentz-equivalent to the spacelike compactification on a circle of
vanishing radius and, in this limit, certain quantities do indeed blow
up.  We have shown that the superpartners of the $r \ra 0$ limit of
the D0-brane possess neither spin nor D2-brane
electric dipole moments.  Magnetic dipole moments, with respect to the
R-R 1-form and NS-NS 2-form, do persist however.  It would be
interesting to understand these results from the perspective of Matrix
theory.

Of course, one could further consider higher order supersymmetry
transformations to determine, for example, the quadrupole moments of
the eleven-dimensional supergraviton.  Via dimensional reduction, this would be a simpler way of
determining the higher order variations of the D0-brane solution,
rather than working with type IIA supergravity directly.  However, the
necessity of compensating Lorentz transformations in eleven dimensions
ensures that such a calculation would still be rather tedious.  

The couplings of the background supergravity fields to the ``worldvolume'' fermions
of the eleven-dimensional supergraviton could also be determined.  And
this would allow for the analysis of interactions between
supergravitons, via the consideration of a probe action for
a massless spinning particle in a general eleven-dimensional
background.  Indeed, using techniques similar to those of~\cite{MTR},
this procedure has been carried out in~\cite{HKS2}, and the resulting couplings have a form
similar to that for the D0-brane in (\ref{eqn:action}).  However, only the
first two terms of (\ref{eqn:action}), coming from the eleven-dimensional
coupling $\del_j h_{+i} \bar{\th} \ga^{ij} \th$, were found
in~\cite{HKS2}.  We have shown that there should be further
couplings of the pp-wave to the background 3-form, and it would be of
interest to determine these.  Of course, they should be of the form
$\del_k A_{+ij} \bar{\th} \ga^{ijk} \th$~\cite{MTR}, but the
explicit computation would provide a nice consistency check.

\acknowledgments
We thank Mark van Raamsdonk for pointing out a misprint
in eqn. (12) of~\cite{MTR}.  DB thanks Carlos Herdeiro, Bert Janssen,
Simon Ross and Paul Saffin for useful conversations and
comments on early drafts of this paper, and is supported in part by
the EPSRC grant GR/N34840/01.  AC thanks Wati Taylor for numerous useful
conversations, and is supported in part by funds provided by the
U.S. Department of Energy (D.O.E.) under
cooperative research agreement DE-FC02-94ER40818.

\bigskip

\appendix

\section{Conventions}

We use the signature $(- + \ldots +)$.  Eleven-dimensional coordinates
$x^A, A, B = 0, \ldots, 10$ are written in terms
of ten-dimensional coordinates $x^a, a,b = 0, \ldots, 9$ and
$z$ (or $x^-$).  The ten-dimensional coordinates are written in terms
of $t$ (or $x^+$) and $x^i, i,j = 1, \ldots, 9$.  Tangent space directions are denoted by an
underline.  The eleven-dimensional Planck
length $l_p = g_s^{1/3} \sqrt{\al'}$, so that $\ka_{11}^2 = 2\pi R
g_s^2 \ka_{10}^2 $ where $g_s$ is the string
coupling constant and the radius of the eleventh dimension is $R = g_s
\sqrt{\al'}$.

The Dirac matrices satisfy
\be
\{\Ga^{\ul{A}}, \Ga^{\ul{B}}\} = 2 \eta^{\ul{A}\,\ul{B}}.
\ee

\noindent Explicitly, the $32 \times 32$ component Dirac
matrices are given by
\be
\Ga^{\ul{i}} = \left( \begin{array}{cc} 0 & \s^{\ul{i}} \\ \s^{\ul{i}} & 0
\end{array} \right), \qquad \Ga^{\ul{t}} = \left( \begin{array}{cc} 0 & -1 \\ 1 & 0
\end{array} \right), \qquad \Ga^{\ul{z}} = \left( \begin{array}{cc} 1 & 0 \\ 0 & -1
\end{array} \right),
\ee

\noindent where $\s^{\ul{i}}$ denote $16 \times 16$ real,
symmetric $SO(9)$ Dirac matrices.  Eleven-dimensional spinors are
anticommuting and Majorana; they obey the useful identities
\be
\bar{\psi}~ \Ga^{\ul{i}_1} \ldots \Ga^{\ul{i}_n} \Ga^{\ul{t}}~ \la = -
\bar{\la}~ \Ga^{\ul{i}_n} \ldots \Ga^{\ul{i}_1} \Ga^{\ul{t}}~ \psi,
\label{eqn:maj1}
\ee
\be
\bar{\psi}~ \Ga^{\ul{i}_1} \ldots \Ga^{\ul{i}_n}~ \la = (-)^n
\bar{\la}~ \Ga^{\ul{i}_n} \ldots \Ga^{\ul{i}_1}~ \psi.
\label{eqn:maj2}
\ee

\noindent Such a spinor can be split as $\ep = P_+ \ep + P_- \ep
\equiv \ep_+ + \ep_-$, where the projection operators
\be
P_{\pm} = \half \left( 1 \pm \Ga^{\ul{t}} \,\Ga^{\ul{z}} \right) =
\half \left( \begin{array}{cc} 1 & \pm 1 \\ \pm 1 & 1 \end{array} \right).
\ee

\noindent In terms of a 16-component spinor $\ep$,
\be
\ep_{\pm} = \left( \begin{array}{c} \ep \\ \pm \ep \end{array} \right).
\ee

\noindent The zero mode fermions in the text are given in terms of
$\ep_+$, and the only non-vanishing fermion bilinears that can be
constructed with this are
\be
\bar{\ep}_+ \,\Ga^{\ul{i}\,\ul{j}\,\ul{t}} \,\ep_+, \qquad \bar{\ep}_+
\,\Ga^{\ul{i}\,\ul{j}\,\ul{k}\,\ul{t}} \,\ep_+.
\ee

\end{document}